\documentclass[aps,prl,twocolumn,showpacs,superscriptaddress]{revtex4-1}  
\usepackage{graphicx}  
\usepackage{dcolumn}   
\usepackage{bm}        
\usepackage{amssymb}   

\begin{document}

\title{Optically addressable silicon vacancy-related spin centers in rhombic silicon carbide with high breakdown characteristics and ENDOR evidence of their structure}
\author{V.A.\,Soltamov}
\email{victorsol85@gmail.com}
\affiliation{Ioffe Institute, St. Petersburg, 194021, Russia}

\author{B.V.\,Yavkin}
\affiliation{Kazan Federal University, Kazan, 420008 Russia}

\author{D.O.\,Tolmachev}
\affiliation{Ioffe Institute, St. Petersburg, 194021, Russia}

\author{R.A.\,Babunts}
\affiliation{Ioffe Institute, St. Petersburg, 194021, Russia}

\author{A.G.\,Badalyan}
\affiliation{Ioffe Institute, St. Petersburg, 194021, Russia}

\author{V.Yu.\,Davydov}
\affiliation{Ioffe Institute, St. Petersburg, 194021, Russia}

\author{E. N. Mokhov}
\affiliation{Ioffe Institute, St. Petersburg, 194021, Russia}

\author{I.I.\,Proskuryakov}
\affiliation{Institute of Basic Biological Problems RAS, Pushchino, 142290 Russia}

\author{S.B.\,Orlinskii}
\affiliation{Kazan Federal University, Kazan, 420008 Russia}

\author{P.G.\,Baranov}
\affiliation{Ioffe Institute, St. Petersburg, 194021, Russia}

\begin{abstract}
We discovered uniaxial oriented centers in silicon carbide having unusual performance. Here we demonstrate that the family of silicon-vacancy related centers with $\textsl{S}$= 3/2 in rhombic 15R-SiC crystalline matrix possess unique characteristics such as optical spin alignment existing at temperatures up to 250$^{\circ}$C. Thus the concept of optically addressable silicon vacancy related centers with half integer ground spin state is extended to the wide class of  SiC rhombic polytypes. The structure of these centers, which is a fundamental problem for quantum applications, has been established using high frequency ENDOR. It has been shown that a family of silicon-vacancy related
centers is a negatively charged silicon vacancy in the paramagnetic state with the
spin $\textsl{S}$= 3/2, V$\tiny_\textrm{Si}$$^{-}$, perturbed by neutral carbon vacancy in non-paramagnetic state, V$\tiny_\textrm{C}$$^{0}$, having no covalent bond with the silicon vacancy and located adjacently to the silicon vacancy on the c crystal axis.
\end{abstract}

\maketitle
Spin centers in Silicon Carbide (SiC) have recently been put forward as favorable candidates for a new generation quantum spintronics and sensorics, as well as quantum information processing
because of the unique properties of their electron spins which can be optically
polarized and read out \cite{bb1,b1,b10,b2,b3,b4,b5,b6,b7,b8,b9}. Particularly the optical control of the single
defect spin in SiC has been realized at room temperature for the first time on the well studied  silicon vacancy-related center with ground spin state $\textsl{S}$= 3/2  in 4H-SiC \cite{b8}. The centers of the same origin persist also in 6H-SiC polytype and were proposed for use in quantum magnetometry \cite{b4} and room temperature operated MASERs \cite{b6}. Thus the family of the V$\tiny_\textrm{Si}$ related centers can be considered to form the basis of the SiC-based quantum sensors and quantum spintronics. The first solid state system on which optical detection and manipulation of the single spin has been shown is the negatively charged nitrogen-vacancy (NV$^{-}$) center in diamond. There are two main reasons why successful realization of single spin control in this system was possible. First,  the  NV$^{-}$ centers possess high optically detected magnetic resonance (ODMR) contrast and their spin alignment persists at elevated temperatures \cite{b11}, thus the polarized spin state can be readout with high fidelity even on the single spin \cite{b12}.  Second,  the precise atomic structure of the center has been elucidated \cite{b13}\cite{b14}. This gave rise to the precise technology of NV's production in predetermined topology and concentrations \cite{b15},\cite{b16}, allowing one to organize precise control of
the NV's electron spin interactions with the environment. This permits to develop NV$^{-}$ based quantum registers \cite{b17}, \cite{b18} and quantum sensors \cite{b19}, \cite{b20}.

Therefore for the successful development of the SiC two major issues have to be addressed. On the one hand spin centers with high breakdown characteristics and high ODMR contrast must be explored, on the other hand the proper model for these spin centers must be established. We focused our efforts on these two tasks and present our results in this work.
  


SiC single crystals of
15R polytype, were grown by the Physical Vapor
Transport method. The concentration of uncompensated nitrogen was in the
range $\approx$10$^{16}$ cm$^{-3}$.  Vacancies
were introduced by irradiation of the crystal with 1.4 MeV electrons with a dose of 10$^{18}$cm$^{-2}$. The crystal structure was confirmed by means of  Raman scattering spectroscopy \cite{b21}, \cite{b22}.
Results are summarized in Supplementary Information (SI). 

\begin{figure}[htbp]%
\includegraphics[scale=1]{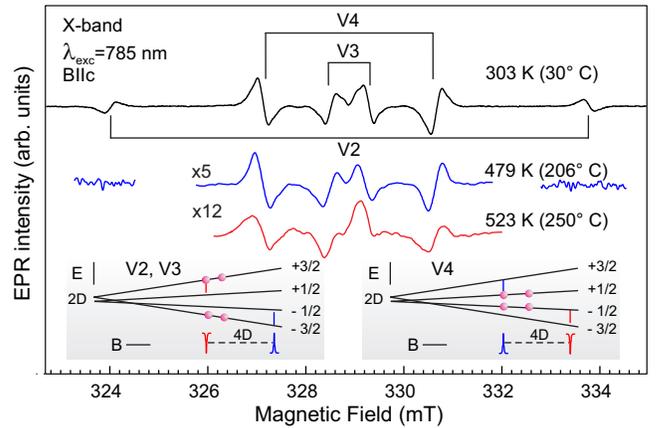}%
\caption{
(Color online) EPR spectra of the
15R-SiC taken at temperatures T= 303K, T= 479K, T= 523K. Insets show light-induced inverse
population of the spin sublevels of V2,V3 and V4.
}%
\label{fig01}%
\end{figure}

First we measured a cw X-band ($\approx$ 9.3GHz) electron paramagnetic resonance (EPR). Fig.1 shows spectra recorded at three different
temperatures (up to 250$^{\circ}$C) under continuous laser light
illumination with $\lambda$= 785nm and the magnetic field oriented parallel to the $\emph{c -}$
axis of the crystal (B$||$$\emph{c}$) (without optical excitation the signals at
room temperature are near the noise threshold). The spectra consist of three
pairs of EPR transitions V2, V3, V4 attributed so from the 
corresponding Zero Phonon luminescence Lines (ZPL) as will be shown further. The
angular variation of the splitting value between each pair of lines (see SI) can be
fitted by the axially symmetric spin Hamiltonian with parameters
summarized in Table I: 
\begin{equation} 
H= g\mu_{B}\vec{B}_{0}\cdot \vec{S} + D(S_Z^{2} - \frac{1}{3}S(S+1))
\end{equation}
where the first and the second
terms correspond to the Zeeman interaction and fine structure splitting,
respectively, $\mu$$_{B}$ and \emph{g} is the Bohr magneton and electron g factor, respectively, total electron spin is $\textsl{S}$= 3/2
and \emph{D} is the fine structure parameter,
e.g., zero-field splitting (ZFS) = 2\textit{D}. The evidence of the $\textsl{S}$= 3/2 state
will be provided from our ODMR  measurements. Thus, all three centers are \emph{c}-axis oriented with C$_{3v}$ point-group symmetry.

\begin{table}
\caption{The values of \textit{D}, \textit{g} , wavelength of ZPLs and T$_1$ and T$_2$ ESE measured at room temperature }
\begin{tabular}{|c|c|c|c|c|c|}

\hline
center	& $\textit{D}$(MHz)	& $\textit{g}$	& ZPL (nm) & $\textsl{T}$$_{1}$ $\mu$s & $\textsl{T}$$_{2}$ $\mu$s \\ \hline
V2	    &69.6$\pm$1	&2.005(1)	&886.5 &   80 &  10  \\ \hline
V3	    &5.8$\pm$1	&2.005(3)	&904   &  120 &  10  \\ \hline
V4	    &25.1$\pm$1	&2.005(3)	&917   &  105 &  9 \\ \hline

\end{tabular}
\end{table}

A phase reversal was observed for one of the two
transitions in each pair of lines. Such behavior in the emission/absorption
mode of the microwave power can be explained by optically induced spin
alignment of the spin sublevels of the centers. Note the
unique properties of these centers, where optical polarization can
be created at high temperatures up to (T$\approx$
250$^{\circ}$C), which indicates that the spin system of the centers is well isolated from the crystalline lattice vibration. In this aspect they are similar to
the NV$^{-}$ centers in diamond, with important
exception --- the ZFS of each center stays stable at the whole temperature range used while the NV-centers experience the huge thermal
shift \cite{b11}, \cite{b23}, \cite{b24}. 

To establish the model of the optical polarization of the V2, V3 and V4 centers we determined the ordering of their spin sublevels (the sign of the ZFS).
Low temperature EPR measurements, conducted at high frequency (95GHz), and Electron Nuclear Double Resonance (ENDOR) study allowed us to determine that ZFS is positive for each center (See SI) and establish the  scheme of the optically polarized spin sublevels as shown in the insets of Fig.1.

We then decided to probe the coherence times of the ensembles of the spin centers using EPR X-band pulsed technique. The results are summarized in Table I for each center. The spin lattice relaxation times ($\textsl{T}$$_{1}$)  and spin-spin relaxation times ($\textsl{T}$$_{2}$) (see Table~1) turned out to be  of the same order of magnitude with those of 4H-SiC spin centers measured previously on the single silicon vacancy related centers \cite{b8}.


\begin{figure}[htbp]%
\includegraphics[scale=1]{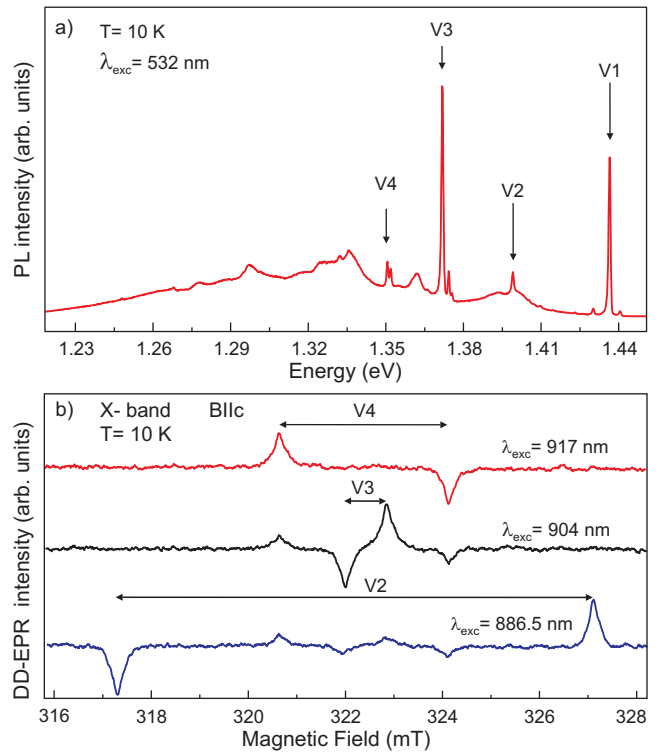}%
\caption{
(Color online) (a) PL spectrum of the 15R-SiC. ZPLs are labeled as
V1, V2, V3, V4. (b) cw X-band DD-EPR spectra obtained at temperature T= 10K and (B$||$$\emph{c}$).}%
\label{fig02}%
\end{figure}

We then proceeded to measure optical properties of the centers discussed above in order to obtain the resonant optical control of their spin states. For such investigations the low temperature photoluminescence (PL) spectrum was measured (Fig.2 (a)). To find the zero photon line (ZPL) that corresponds to each of the EPR centers shown on the Fig. 1 we excited each ZPL with laser flash.  Excitation of each ZPL gives rise to one pair of EPR lines with ZFS corresponding to those determined from our cw-EPR measurements (Table 1). Resulting spectra are summarised in Fig.2 (b) and Table 1. Particularly excitation into V4 line with energy E= 1.3522 eV induced the signal with ZFS equal to 50 MHz, in V3 (E= 1.3716 eV) and V2 (E= 1.3989 eV) with ZFS 13 MHz and 138 MHz, respectively. So we have shown that each center in 15R-SiC has its own optical fingerprint. Excitation into V1 line didn’t give rise to any EPR signal. 

For any application in spin center-based spintronics or quantum sensorics
one have to get access to effective readout of polarized spin state. 
We further demonstrate that V2, V3, V4 centers can be optically
addressed and readout with high fidelity at room temperature by means of a
standard ODMR technique.

\begin{figure}[htbp]%
\includegraphics[scale=1]{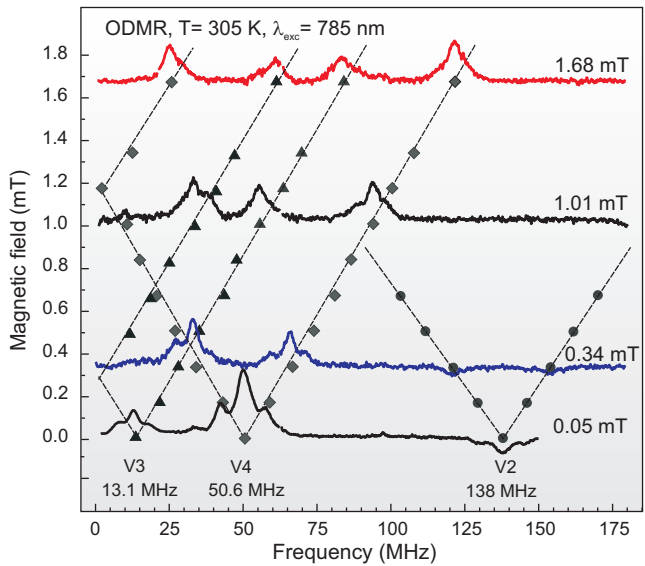}%
\caption{
(Color online) Lower spectrum shows ODMR signal of the V2, V3, V4 detected under 
laser excitation $\lambda$= 785 nm in 0.05 mT external magnetic field applied 
in order to compensate the influence of Earth magnetic field. 
Experimental frequency dependence of the ODMR signals as a function of
magnetic field are shown for each center. Dashed lines are the calculated dependence for the allowed
transitions using equation (1), $\textsl{S}$= 3/2
}
\label{fig03}%
\end{figure}

The ODMR spectrum in the Figure 3
shows relative change of the photoluminescence intensity $\Delta$PL/PL as a
function of applied RF. Resonances at $\nu = 13.1$~MHz,
$\nu = 50.6$~MHz and $\nu = 138$~MHz were observed that agree well with ZFS of the
centers V3, V4 and V2 determined from our EPR measurements.

Using spin Hamiltonian from the Eq. (1) we further calculated that the ground states of the V2, V3 and V4 have a spin state $\textsl{S}$= 3/2. Calculated frequency dependencies of the ODMR signals at different magnetic fields (B$||$$\emph{c}$) coincided well with observed spectra (Figure 3). 

The unusual properties of spin centers in 15R-SiC  observed in this work are several times higher compared to similar signals in the 6H- and 
4-H polytypes. Taking into account high temperatures at which the optical polarization persisted (up to 250$^{\circ}$C) and strong ODMR contrast these centers seem to be favorable candidates for the spin center-based quantum applications. 
Moreover,  changing the excitation energy and the optical registration window, the ODMR contrast can be further increased. When ODMR signal was measured in 15R-SiC under excitation of the V4 center with laser $\lambda$= 808 nm  we observed about five times more intense signals than in other polytypes (unpublished results).  

From the experiments described above it becomes clear that the V$\tiny_\textrm{Si}$ related centers with optically addressable $\textsl{S}$= 3/2 ground states exist not only in hexagonal but also in rhombic polytypes. 
In order to advance understanding functionality of the observed spin systems we determined the microscopic model of the centers.
Currently there are two such models under consideration. The first model postulated these centers to be the low symmetry configuration of the V$\tiny_\textrm{Si}$$^{-}$ with $\textsl{S}$= 3/2 (V$\tiny_\textrm{Si}$$^{-}$) \cite{b27}, \cite{b28}. According to the second model these centers are the V$\tiny_\textrm{Si}$$^{-}$ $\textsl{S}$= 3/2, perturbed by neutral carbon vacancies in nonparamagnetic state, V$\tiny_\textrm{C}$$^{0}$, located adjacently to the silicon vacancy V$\tiny_\textrm{Si}$$^{-}$ on the  $\textit{c -}$ axis of the crystal, see \cite{b4} and references therein.

In this work we present the strong evidence in
favor of the second model by means of Electron Spin Echo (ESE) detected high frequency ENDOR.

To this end we measured the W-band (95 GHz) EPR spectra of
the V2, V3 and V4 centers under $\lambda$= 785 nm
laser excitation for two orientations ($\theta$) of the
magnetic field \emph{B} with the respect to
the \emph{c} axis (Fig.4 (a)). We here focused on the V2 center with larger ZFS= 138 MHz to obtain its ENDOR spectra. Allowed dipole magnetic transitions with $\Delta$\emph{M$_{S}$}=
$\pm$1: $3/2 \leftrightarrow 1/2$ are
indicated by abbreviation lf (low field), $-3/2 \leftrightarrow
-1/2$ indicated by hf (high field). 
The ENDOR
spectra were obtained by monitoring the intensity of the ESE, following three
microwave $\pi$/2
pulses, as a frequency function of the pulse, applied 
between the second and third microwave pulses \cite{b29} (Fig.4 (b,c)). To
describe the hyperfine (HF) interaction with nuclei of silicon and carbon,
 which reside in different shells, one should add to the Spin Hamiltonian given in
Eq. (1) the term: 

$\Sigma$\textbf{SA}$_{i}$\textbf{I}$_{i}$
+ $\Sigma$\textbf{SA}$_{j}$\textbf{I}$_{j}$
+ $\Sigma$\textbf{SA}$_{k}$\textbf{I}$_{k}$,
where \textbf{A}$_{i}$ and \textbf{A}$_{k}$
are tensors which describe the HF interaction with i-th Si and k-th C atoms located in different neighbor shells of the Si sites,
\textbf{A}$_{j}$ is tensor that describes the HF interaction  with j-th Si atoms located in neighbor shell to the C site.

The ENDOR transition frequencies determined by the selection rules $\Delta$\emph{M$_{S}$}=0 and $\Delta$\emph{M$_{I}$} =$\pm$1 are
given by \cite{b30}
\begin{equation}
\nu_{ENDORi} = h^{-1}|\emph{M$_{S}$}(a_{i} + b_{i}(cos^{2}\theta- 1))-g_{ni}\mu_n B| 
\end{equation}
Here
\emph{a$_{i}$}
and \emph{b$_{i}$ }denote the
isotropic and anisotropic part of the interaction with the \emph{i}-th nucleus (e.g.,
for Si), respectively, \emph{$\theta$ }is the
angle between the direction of the external magnetic field \emph{B} and the hyperfine interaction tensor, \emph{g$_{ni}$}$\mu$\emph{$_{ni}$}\emph{B}/\emph{h}
is the nuclear Zeeman term or the Larmor frequency f$_{L}$, \emph{g$_{ni}$} and $\mu$\emph{$_{ni}$} are g factor of nucleus \emph{i} and its Bohr magneton, respectively (\emph{g$_{n}$}
is negative for $^{29}$Si and positive for $^{13}$C). For axial
symmetry HF interaction in terms of the principal values are given by A$_{||}$
= \emph{a} + \emph{b}, A$_{\bot}$
= \emph{a} - \emph{b}.

\begin{figure}[htbp]%
\includegraphics[scale=1]{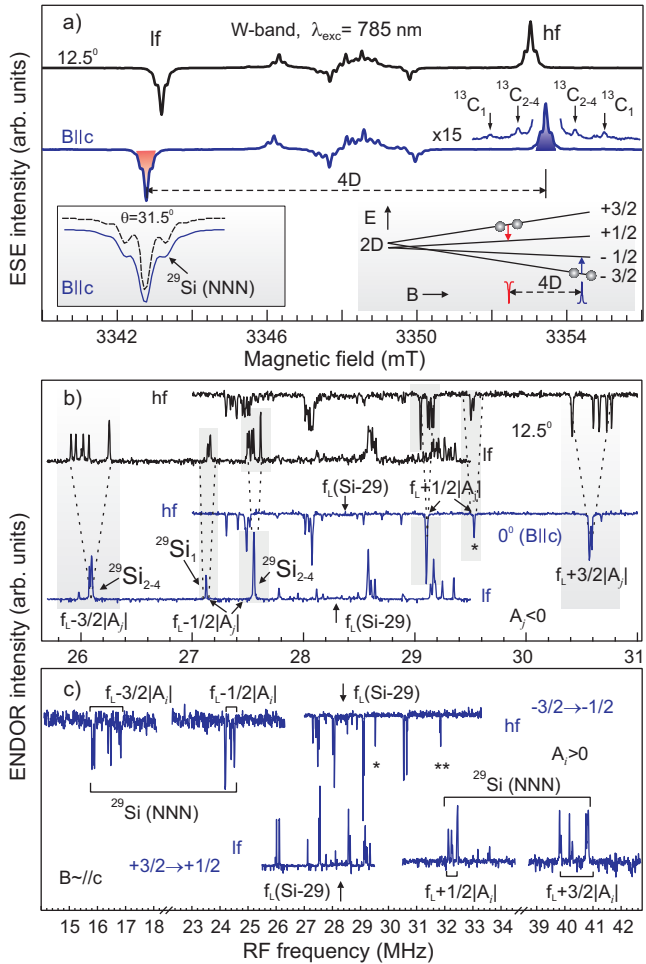}%
\caption{
(a) (Color online)  Optical induced ESE spectra obtained in the
15R-SiC at two angles ($\theta$): $\theta$ = 0$^{0}$ and
$\theta$= 12,5$^{0}$. 
(b), (c) ENDOR spectra measured at the lf and hf transitions indicated in (a). 
$^{29}$Si$_{1}$ and $^{29}$Si$_{2-4}$$_{
}$indicate ENDOR lines correspond to the presence of $^{29}$Si atoms
in the NN shell with respect to V$\tiny_\textrm{C}$$^{0}$. One and two asterisks in (b)
show transitions corresponding to f$_{L}$ + 1/2$|$A$_{j}$$|$ and f$_{L}$
+ 3/2$|$A$_{j}$$|$ for HF interactions with axial NN Si atom with respect to V$\tiny_\textrm{C}$$^{0}$.
}%
\label{fig04}%
\end{figure}

Our data suggest the structural model of the V2 centers as shown in Fig.5. The model represents two possible configurations (V$\tiny_\textrm{Si}$$^{-}$ -- V$\tiny_\textrm{C}$$^{0}$)$^{'}$ and (V$\tiny_\textrm{Si}$$^{-}$ -- V$\tiny_\textrm{C}$$^{0}$)$^{''}$. First, prior to the analysis of obtained ENDOR data, we would like to discuss the HF interactions directly observable from the ESE spectra in Figure 4 (a).

The hyperfine structure,shown in the enlarged scale on the left inset of Fig.5(a), arises from the HF interactions with $^{29}$Si in the next nearest neighbor (NNN) shell to the V$\tiny_\textrm{Si}$$^{-}$ site,  and is characterized by the  constant A= 2.97G. The similar interaction was observed by others in 4H- and 6H-SiC \cite{b31,b26}. Note that changing the orientation of the magnetic field does not induce the line shifts; meanwhile the strong anisotropy of the ESE line width was observed. 
On the other hand, the HF interactions with nucleus located in the first NN shell to the V$\tiny_\textrm{Si}$$^{-}$ site are known to be strongly anisotropic and reflect the tetrahedral symmetry of the nuclear spin locations.  Hyperfine lines arising from interactions with $^{13}$C  are shown in Fig.4 (a). $^{13}$C$_{1}$ denotes interaction with the carbon atom oriented along the $\textit{c -}$ axis and the $^{13}$C$_{2-4}$ denotes the interactions with carbon atoms located in the basal plane with bonds inclined by the angle $\theta$= 71${^0}$ relatively to the $\textit{c -}$ axis. These carbon atoms are shown schematically in Fig.5. The HF structure arising from such interactions is characterized by the parameters A$_{||}$ = 30.2 G (84.6 MHz) and A$_{\bot}$ = 12 G (33.6 MHz) which match closely previously reported values for the V$\tiny_\textrm{Si}$$^{-}$ centers in 4H- and 6H-SiC 
\cite{b26,b31,b32,b33}.
Therefore our ESE measurements point to  strongly anisotropic nature of HF interactions of the V$\tiny_\textrm{Si}$$^{-}$ electron spin with carbon nuclear spins in the NN shell and reflect their tetrahedral configuration. 

From the measured ENDOR spectra (Fig.4 (b,c)) all the lines observed can be attributed to the HF interactions of V$\tiny_\textrm{Si}$$^{-}$ with $^{29}$Si. To reveal this we need to consider Eq.2, which predicts that for $\textsl{S}$= 3/2  each nuclei i induces two sets of ENDOR transitions located at a distance 3/2A$_{i}$ and 1/2A$_{i}$ for the lf EPR line, -3/2A$_{i}$  and -1/2A$_{i}$ for the hf, from the corresponding Larmor frequency f$_{L}$. Thus, the ENDOR lines observed are due to HF interactions with $^{29}$Si, because the f$_{L}$ of the $^{29}$Si (I= 1/2, abundance 4.7\%) is about 28.2MHz. The more accurate value of the f$_{L}$ depends on the magnitude of the magnetic field for the EPR transition and is marked by arrows in Fig. 4(b,c). 
HF interactions of the V$\tiny_\textrm{Si}$$^{-}$ electron spin with $^{29}$Si nuclear magnetic moments in the NNN shell observed on the ESE spectra can also be seen in our ENDOR spectra (Fig. 4 (c), labeled  $^{29}$Si(NNN)). Position of these lines corresponds to  the positive density of the electronic wave function around the silicon vacancy. Unexpectedly we also observed the lines that  were characterized by the negative spin density on Si nucleus.
 In Fig. 4(b) we show the expanded part of the ENDOR spectra for lines that correspond to the negative spin density  measured at two different orientations of B relative to the $\textit{c -}$ axis ($\theta$= 0${^0}$ and $\theta$= 12.5${^0}$). Strong anisotropy of the signals (shown by shadowed area in Fig.4(b)) was observed.  Earlier in discussion of ESE spectra we showed that interactions with $^{29}$Si located in NNN and more distant shells are almost isotropic, meanwhile, interactions with $^{13}$C located in the NN shell possess anisotropy. In ENDOR spectra we observed for HF interactions with with negative spin density $^{29}$Si the same anisotropic dependence as that for $^{13}$C located in the NN shell seen in ESE. To explain such an anisotropy, one has to identify in the model (Fig. 5) Si atoms which reflect the same symmetry as that for C atoms in NN shell of silicon vacancy. For Si atoms, such a configuration could be found only at the tetrahedron vertices around the carbon site in the SiC lattice. Position of ENDOR lines labeled as $^{29}$Si$_{1}$ and $^{29}$Si$_{2-4}$ in Figure 4(b) agrees well with proposed configuration and reflects the HF interactions with axial (Si$_{1}$) and basal  (Si$_{2-4}$)  nuclear spins. The line corresponding to HF interactions with axially oriented Si nuclei for transitions with \emph{M$_{S}$}=±3/2 is shown on Fig.4 (c) marked by two asterisks. The constants of HF interactions with negative spin density are relatively large and correspond approximately to the A$_{||}$ = 2.2 MHz ($\approx$0.8 G) и A$_{\bot}$ =1.3 MHz ($\approx$~0.5 G). Notably these constants describe well the anisotropy of the linewidth observed on ESE. 

HF interactions with negative spin density $^{29}$Si could be explained if the spin density was located on four Si nuclei placed around the carbon vacancy being in the non-paramagnetic neutral charge state, V$\tiny_\textrm{C}$$^{0}$. Therefore we conclude that both the V$\tiny_\textrm{C}$$^{0}$ and paramagnetic $\textsl{S}$= 3/2 V$\tiny_\textrm{Si}$$^{-}$ form the V2 center. Negative spin density is caused by the spin polarization (similar to the core polarization for transition metals \cite{b34,b35}). This spin polarisation arises from exchange interaction with $\textsl{S}$= 3/2 that leads to partial decoupling of coupled covalent bonds in the V$\tiny_\textrm{C}$$^{0}$ site. The other interactions with smaller negative spin density observed on ENDOR spectra could be attributed to HF interaction with more distant Si nuclei from the V$\tiny_\textrm{C}$$^{0}$ site. 
V$\tiny_\textrm{C}$$^{0}$ distorts the crystal lattice which in turn lowers symmetry of the V$\tiny_\textrm{Si}$$^{-}$, so the carbon vacancy could be expected at the center with the largest ZFS (V2) in the position most close to the V$\tiny_\textrm{Si}^{-}$.
It is important to note that the double number of lines observed in the ENDOR spectra  points to the presence of two similar centers with slightly different parameters of HF interactions (Fig.4 b, $\theta=12.5^{\circ}$). In 15R-SiC it could be connected with the unit cell structure. Thus, the model of the V2 center could be reconstructed as shown in the Fig.5. 
The ENDOR spectra of (V3) and (V4) in 15R-SiC (not shown here) exhibited the same pattern of the HF interactions with $^{29}$Si nuclei, having slightly different values of hyperfine splitting, so the structure of these centers has to be similar to that of V2 (V$\tiny_\textrm{Si}$$^{-}$ -- V$\tiny_\textrm{C}$$^{0}$). The only difference would be that the distance between carbon-silicon vacancies along the $\textit{c -}$ axis should be larger than that of V2. 
Moreover, we have shown here that the properties of V$\tiny_\textrm{Si}$$^{-}$ related centers with $\textsl{S}$= 3/2 in 15R SiC resemble those of 4H, 6H , therefore the model proposed here for V2 in 15R (Fig. 5) could be extended for hexagonal polytypes. Supporting evidence for possible model universality will be provided in later publications.
\begin{figure}[htbp]%
\includegraphics[scale =1]{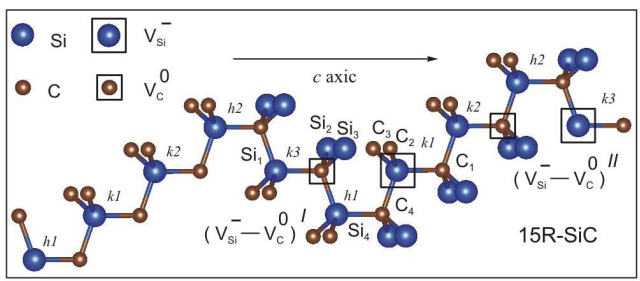}%

\caption{
(Color online) The fragment of the 15R-SiC structure with two hexgonal (\textit{h1}, \textit{h2})
and three quasicubic (\textit{k1}, \textit{k2} and \textit{k3}) sites for silicon. Vacancies are labeled by squares. Carbon and Silicon atoms
in the NN shell to the V$\tiny_\textrm{Si}$$^{-}$ and V$\tiny_\textrm{C}$$^{0}$
are labeled as C$_{1-4}$ and Si$_{1-4}$, respectively. Structure of the V2 center is indicated as (V$\tiny_\textrm{Si}$$^{-}$ -- V$\tiny_\textrm{C}$$^{0}$)$^{'}$ and (V$\tiny_\textrm{Si}$$^{-}$ -- V$\tiny_\textrm{C}$$^{0}$)$^{''}$.
}%
\label{fig05}%
\end{figure}

 In conclusion in this work we have provided evidence for the presence of three uniaxial spin centers in 15R-SiC lattice. These centers with distinct ZPL and ZFS were characterized by long spin coherence times at room temperature.  Their optical spin alignment persisted up to 250$^{\circ}$C and the value of  the ZFS did not show temperature dependence. ENDOR study revealed the presence of two types of HF interactions of the electron spin of these centers  with surrounding $^{29}$Si nuclear spins, characterized by negative and positive spin densities.  This allowed us to establish the model for the V2 center.  
This knowledge is important for understanding the processes leading to the decoherence \cite{b36} and paves the way to the effective and precise production of silicon vacancy related centers.
 The newly discovered silicon vacancy-related centers are well isolated from the SiC crystal matrix, therefore these centers lend themselves for further development of SiC based quantum applications and quantum registers based on coupled electron spins.
Furthermore, our results extend the concept of optically addressable V$\tiny_\textrm{Si}$$^{-}$ related centers to the rhombic polytypes of silicon carbide.

\begin{acknowledgments}
This work has been supported by MES of Russia No~14.604.21.0083 and by the RFBR No~13-02-00821.

\end{acknowledgments}


\begin{thebibliography}{}


\bibitem{bb1}
P. G. Baranov, A. P. Bundakova, I. V. Borovykh, S. B. Orlinski, R. Zondervan, J. Schmidt,
JETP Letters \textbf{86}, 202–206 (2007);
\bibitem{b1}
P. G. Baranov, A. P.
Bundakova, A. A. Soltamova,
 S. B. Orlinskii, I. V. Borovykh,R. Zondervan, R. Verberk,J. Schmidt, Phys.
Rev. B \textbf{83}, 125203 (2011).
\bibitem{b2} 
V. A. Soltamov, A. A. Soltamova, P. G. Baranov, I. I. Proskuryakov, Phys. Rev. Lett. \textbf{108}, 226402 (2012).
\bibitem{b10}
W. F. Koehl, B. B. Buckley, F. J. Heremans, G. Calusine, and D. D. Awschalom,
Nature (London) \textbf{479}, 84 (2011).
\bibitem{b3} 
A. L. Falk, B. B. Buckley, G. Calusine, W. F. Koehl, V. V. Dobrovitski, A. Politi, C. A. Zorman, P. X.-L.
Feng, and D.D. Awschalom, Nat. Commun.
\textbf{4}, 1819 (2013).
\bibitem{b4} 
H. Kraus, V. A. Soltamov, D. Riedel, S. V{\"a}th, F. Fuchs, A. Sperlich, P. G. Baranov, V. Dyakonov, and G. V. Astakhov, Nature Phys. \textbf{10}, 157-162 (2014).
\bibitem{b5} 
A. L. Falk, P. V. Klimov, B. B. Buckley, V. Iv{\'a}dy, I. A. Abrikosov, G. Calusine, W. F. Koehl, {\'A}. Gali, and D. D. Awschalom, Phys.
Rev. Lett. \textbf{112},
187601 (2014).
\bibitem{b6} 
H. Kraus, V. A. Soltamov, F. Fuchs, D. Simin, A. Sperlich, P. G. Baranov, G. V. Astakhov \& V. Dyakonov,
Sci. Rep. \textbf{4},\textbf{ }5303 (2014).
\bibitem{b7} P. V. Klimov, A. L. Falk, B. B. Buckley, and D. D. Awschalom, Phys. Rev. Lett. \textbf{112}, 087601 (2014).
\bibitem{b8} M. Widmann, S.-Y. Lee, T. Rendler, N. T. Son, H. Fedder,
S. Paik, L.-P. Yang, N. Zhao, S. Yang, I. Booker, A. Denisenko, M. Jamali, S. A. Momenzadeh, I. Gerhardt, T. Ohshima, A. Gali,
E. Janz{\'e}n and J. Wrachtrup, Nat. Mater. \textbf{14}, 164-168 (2015).
\bibitem{b9} 
D. J. Christle, A. L. Falk, P. Andrich, P. V. Klimov, J. Hassan, N. T. Son, E. Janz{\'e}n, T. Ohshima, and D. D. Awschalom,
Nat. Mater. \textbf{14},
160 (2015).

\bibitem{b11}
D. M. Toyli, D. J. Christle, A. Alkauskas, B. B. Buckley, C. G. Van de Walle, and D. D. Awschalom, Phys.
Rev. X \textbf{2}, 031001 (2012).
\bibitem{b12}
A. Gruber, A. Dra{\"u}benstedt, C. Tietz, L. Fleury, J. Wrachtrup, and C. Borczyskowski, Science \textbf{276}, 2012 (1997).
\bibitem{b13}
J.H.N. Loubser, J.A. van Wyk,
Rep. Prog. Phys. \textbf{41} (1978).
\bibitem{b14} D.A.
Redman, S. Brown, R.H. Sands, S.C. Rand, Phys. Rev. Lett. \textbf{67}, 3420 (1991).
\bibitem{b15}
J. Martin, R. Wannemacher, J. Teichert, L. Bischoff, and B. Kohler, Appl. Phys. Lett. \textbf{20}, 3096–3098 (1996).
\bibitem{b16}
Pezzagna S., Wildanger D., Mazarov P., Wieck A. D., Sarov Y., Rangelow I., Naydenov B., Jelezko F., Hell S. W. and Meijer J., Small \textbf{6}, 2117–2121 (2010).
\bibitem{b17} 
G.Waldherr, Y.Wang, S. Zaiser, M. Jamali, T. Schulte-Herbr{\"u}ggen, H. Abe, T. Ohshima, J. Isoya, J. F. Du, P. Neumann \& J. Wrachtrup, Nature \textbf{506},  204 (2014).
\bibitem{b18}
T. H. Taminiau, J. Cramer, T. van der Sar, 
V. V. Dobrovitski and R. Hanson, Nature Nanotech. \textbf{9 }(2014).
\bibitem{b19} 
V. M. Acosta, E. Bauch, M. P. Ledbetter, C. Santori, K.-M. C. Fu, P. E. Barclay, R. G. Beausoleil, H. Linget, J. F. Roch, F. Treussart, S. Chemerisov, W. Gawlik, and D. Budker, Phys. Rev. B \textbf{80}, 115202 (2009).
\bibitem{b20}
F. Dolde, M. W. Doherty, J. Michl, I. Jakobi, B. Naydenov, S. Pezzagna, J. Meijer, P. Neumann, F. Jelezko, N. B. Manson, and J{\"o}rg Wrachtrup,
Phys. Rev. Lett. \textbf{112}, 097603 (2014).
\bibitem{b21} S. Nakashima and H. Harima, Phys. stat. sol. (a) \textbf{162}, 39 (1997).
\bibitem{b22}
D. W. Feldman, J. H. Parker, Jr., W. J. Choyke, and
L. Patrick, Phys. Rev. \textbf{170}, 698
(1968).
\bibitem{b23} V. M. Acosta, E. Bauch, M.
P. Ledbetter, A. Waxman, L.-S. Bouchard,
and D. Budker Phys. Rev. Lett. \textbf{104}, 070801 (2010).
\bibitem{b24}
R. A. Babunts, A. A. Soltamova, D. O. Tolmachev, V. A. Soltamov, A. S. Gurin, A. N. Anisimov, V. L. Preobrazhenskii, P. G. Baranov, JETP Lett. \textbf{95}, 477-480 (2012).
\bibitem{b25}
E. Sorman, N. T. Son, W. M. Chen, O. Kordina, C. Hallin, and E.
Janzen, Phys. Rev. B \textbf{61}, 2613
(2000).
\bibitem{b26}
H. J. von Bardeleben, J. L. Cantin,
I. Vickridge, and G. Battistig,
Phys. Rev. B \textbf{62}, 10126 (2000).
\bibitem{b27}
N. Mizuochi, S. Yamasaki, H. Takizawa, N. Morishita, T. Ohshima, H. Itoh, T. Umeda,
and J. Isoya, Phys. Rev. B \textbf{72}, 235208 (2005). 
\bibitem{b28}
Janz{\'e}n, Erik; et al. Physica
B \textbf{404}, 4354-4358 (2009).
\bibitem{b29} W. B. Mims, in Electron Paramagnetic Resonance,
edited by S. Geschwind, Plenum, New York, (1972).
\bibitem{b30}
J.-M. Spaeth, J.R. Niklas, R.H. Bartram, Structural Analysis of Point Defects in Solids, Chapter 5, p. 152, Springer-Verlag, Berlin, Heidelberg
(1992).
\bibitem{b31}
T. Wimbauer, B.K. Meyer, A. Hofstaetter, A. Scharmann, H. Overhof,
Phys. Rev. B \textbf{56} ,7384 (1997).
 
\bibitem{b32}
N. Mizuochi, S. Yamasaki, H. Takizawa, N. Morishita, T. Ohshima, H. Itoh, and J. Isoya, Phys. Rev
B \textbf{66}, 235202 (2002).
\bibitem{b33}
W. E. Carlos, N. Y. Garces,  E. R. Glaser, M.A. Fanton, Phys. Rev
B \textbf{74}, 235201 (2006).

\bibitem{b34}
A.J. Freeman and R.B. Frankel, Hyperfine Interactions, Academic Press New York
London (1967). 
\bibitem{b35}
A. Abraham and B. Bleaney, Electron Paramagnetic Resonance of Transition Ions, pp.702-705, Clarendon Press, Oxford (1970). 
\bibitem{b36}
Li-Ping Yang, Christian Burk, Matthias Widmann, Sang-Yun Lee, J{\"o}rg Wrachtrup, and Nan Zhao, Phys. Rev. B \textbf{90},
241203(R) (2014).

\end{thebibliography}
\end{document}